# Power Cycling Test Bench for Accelerated Life Testing for Reliability Assessment of SiC-MOSFET in Extreme Offshore Environment

Shiva Geraei(1), Saeed Hasanpour Aghdam(2)
(1) Electrical and Computer Engineering, Concordia University, s_garaei@encs.concordia.ca
(2) Electrical and Computer Engineering, University of Tabriz, Iran – saghdam@tabrizu.ac.ir

## Abstract

The reliability of power semiconductor switches is important when considering their vital role in power electronic converters for downhole subsea applications. Respect to technology advancements in material sciences, power MOSFETs with wide band gap materials have been proposed such as silicon carbide (SiC) and gallium nitride (GaN) as an alternative to existing silicon (Si) based MOSFETs and IGBTs. However, reliability analysis should be performed before substituting SiC-MOSFETs in the place of existing Si-MOSFETs and IGBTs. Due to costly equipment of experimental test setup for accelerated life test, a good reliable and precise simulation-based test bench should be used to test the life test procedure before implementing actual hardware. Therefore, this paper introduces a power cycle (PC) test bench for accelerated life testing for reliability assessment of SiC-MOSFET in harsh offshore environment. The introduced test bench is a simulation-based of power switch in SimScape and LTspice and has been validated with datasheet of 1.2 kV SiC-MOSFET, CAS300M12BM2 by CREE. Preliminary hardware circuits are also shown for further experimental tests. The captured data from the Device-Under-Test (DUT) in different ambient temperatures are envisioned and provide critical information about the failure mechanisms and lifetime characteristics of power devices. The provided lifetime characteristics data of SiC-MOSFET can be used to statistically estimate the Remaining-Useful-Lifetime (RUL) of component in a real application such as downhole motor drives.

## I. Introduction

Power semiconductor devices have become a crucial component in most of the modern energy systems varying from wind turbines to hybrid electric vehicles [1]. Throughout the converter operation, power semiconductor devices may experience several unexpected and abnormal stresses due to environmental conditions, transients and overload conditions, which can lead to severe degradation or even failure and ultimately interrupt the operation of the power converter. In particular, for applications in remote locations such as offshore wind systems and human life risk related applications such as automotive and transportation applications, a clear understanding of the failure modes of the semiconductor devices and prognostic information is absolutely needed for safe, reliable and cost effective operation [2]. Therefore, a better understanding of the failure mechanisms and barriers to the utilization of electronic devices in extreme environments lead to reliable power converters in offshore drilling applications.

According to a survey based on over 200 products from 80 companies, as shown below in Fig. 1, the semiconductor devices and capacitors are the two major components responsible for power electronic systems failure [3]. Therefore, the efficiency and reliability of these components are a great interest as they play a major role in the overall reliability of the power converters.

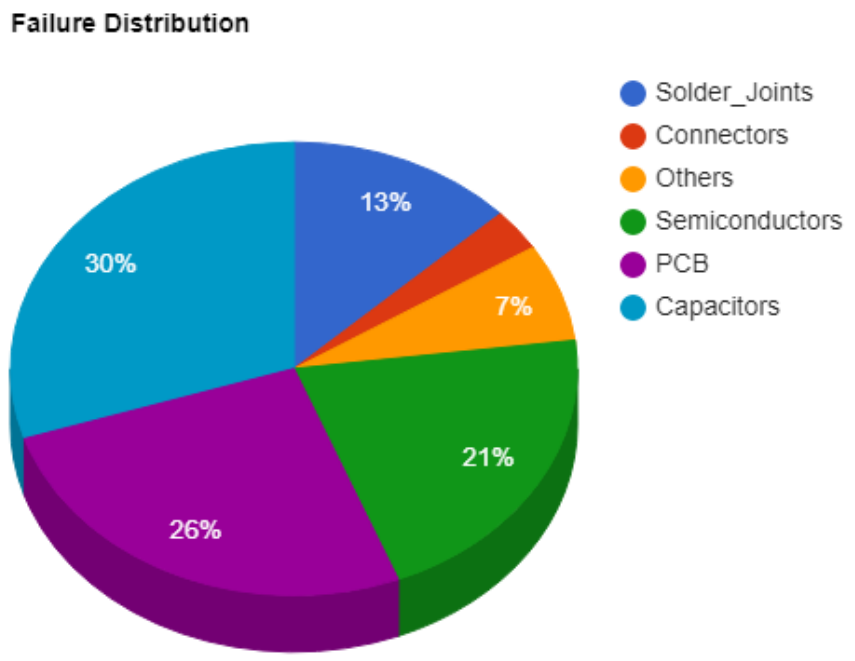

**Fig. 1. Failure distribution in power electronic system [3]**

A better understanding of the failure mechanisms and barriers to the utilization of electronic devices in extreme environments lead to reliable power converters in offshore drilling applications. With the increasing use of oil & gas sources, the number of power converters are increasing. In [4, 5], a novel and fault-tolerant isolated DC/DC converter for offshore subsea application is proposed. Besides, the reliability and efficiency analysis is a great concern and playing significant role in safe operation of power converters in different applications e.g. electric vehicle (EV) and liner motor propulsion control system [6, 7].

Due to continuous emergence of SiC-MOSFETs in power electronic converters, many questions about their performance and reliability rise. The characterization of SiC-MOSFETs were described in [8-11] and their performance is compared to Si-IGBTs and Si-MOSFETs in [12-14].

Since the benefits of the SiC-MOSFETs when compared with conventional Si devices have been presented in the above studies, in this paper a simulation-based characterization of SiC-MOSFET with accelerated life testing platform with repeated high pulse currents and high electric field stress were introduced.

## II. Modeling of SiC-MOSFET

Having an accurate SiC device model can help to design of device driver circuit, protection circuit and EMI suppression method. In recent years, many papers have proposed SiC MOSFET models [15-17] based on PSpice. These models are quite qualified for small circuit simulation, but cannot be used for complex system simulation. Paper [18] proposes a physics-based model of a power SiC BJT, realized using MATLAB/Simulink. A SiC-MOSFET model based on MATLAB/Simulink is described in [19], but the linear region error of output characteristic is difficult to accept.

This paper presents an accurate SiC power MOSFET model based on MATLAB/Simulink and in order to give practical meaning and validation to the obtained results, a commercially available device, 1200V SiC-MOSFET - CAS300M12BM2, has been considered to compare provided characteristics in datasheet with simulation results.

## III. IV-characteristics:

The static characterizations of SiC-MOSFET include DC characteristics (I-V) is described in this section. When the device is operating in the linear region the drain current of the device can be expressed as [20]:

$$I_D = \mu \frac{W_{channel}}{L_{channel}} \frac{C_{ox}}{2} \left[ 2\left(V_{GS} - V_T\right) V_{DS} - V_{DS}{}^2 \right] \tag{1}$$

where $\mu$ is the channel mobility, $W_{channel}$ is the width of the channel, $L_{channel}$ is the length of the channel, and $C_{ox}$ is the per area capacitance of the gate. Equation 1 is usually simplified by assuming the device state. When the drain-source voltage is small the channel is not fully saturated and the device is in its linear mode. When the drain-source voltage is large, the channel is assumed to be fully saturated and an increase in drain-source voltage will not result in a further increase of the drain current. When $V_{DS}$ is small, the MOSFET is the linear region. In this region, the drain-source voltage can be assumed to be small; hence the term $V_{DS}^2$ can be neglected. Therefore, the IV-characteristic in the linear region can be approximated by:

$$I_D = \mu \frac{W_{channel}}{L_{channel}} \left[ \left(V_{GS} - V_T\right) V_{DS} \right] \tag{2}$$

From the equation 2, the equivalent resistance of the device in the linear region can be found as the ratio of change in drain source voltage to change in drain current.

$$R_{ON} = \frac{dV_{DS}}{dI_D} \frac{L_{channel}}{\mu W_{channel} C_{ox} \left(V_{GS} - V_T\right)} \tag{3}$$

From the equation 3, it can be seen that the resistance of the MOSFET depend on its material properties, device geometry, and the applied gate voltage. When the drain source voltage increases, the channel will saturate and the drain current will no longer increase as $V_{DS}$ is increased. The condition were this occurs can be found from the point where the derivative of the drain current with respect to the drain-source voltage is zero. The solution to this equation is the so-called pinch-off condition, which is the voltage at which the MOSFET is saturated. This is given by:

$$V_{DS} = V_{GS} - V_T \tag{4}$$

In the saturation region the current can be found by considering the drain-source voltage to be fixed at the pinch of voltage. Substituting equation 2.4 in equation 2.1 the saturation current can be expressed as:

$$I_D = \mu \frac{W_{channel}}{L_{channel}} \frac{C_{ox}}{2} \left[ \left(V_{GS} - V_T\right)^2 \right] \tag{5}$$

From the equation 5 it can be seen that the current is independent of the drain-source voltage once the channel is fully saturated. Figure 2.a shows the simulation-based output characteristics of a device at different gate voltages. With compare to the manufacturer’s datasheet in Fig. 2.b, it can be seen the accuracy of the modeled SiC-MOSFET.

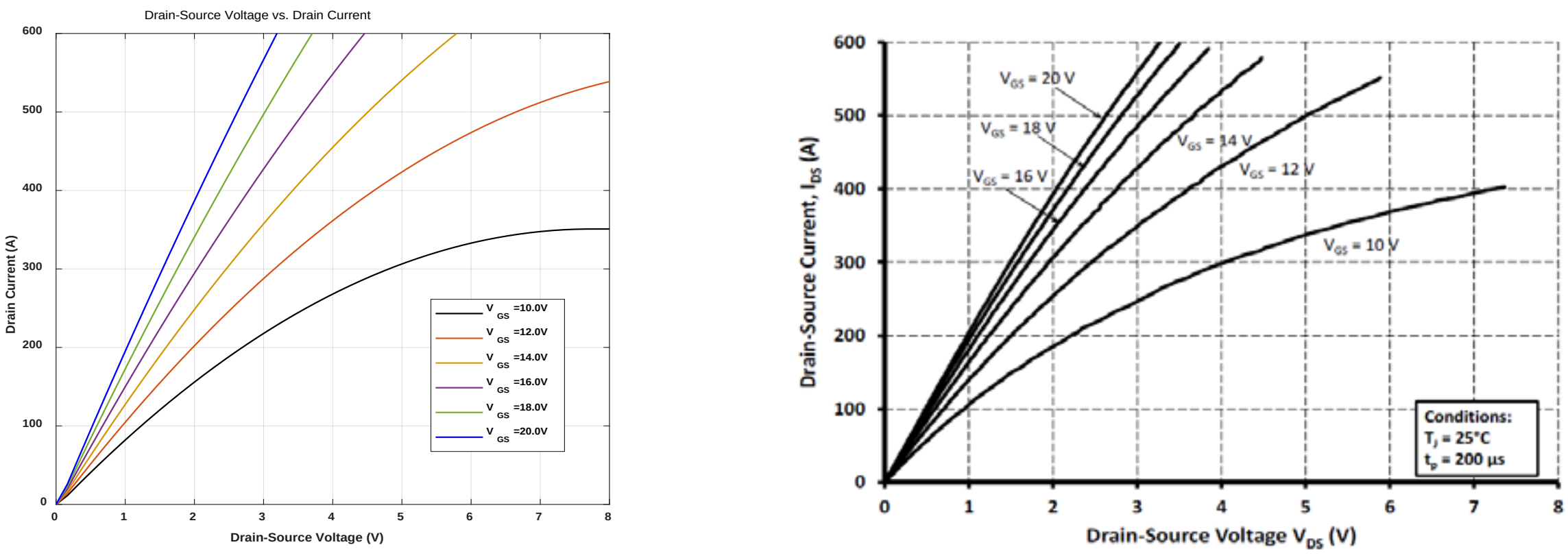

**(a)** **(b)**

**Fig. 2. IV-characteristics of CAS300M12BM2 SiC-MOSFET at different gate voltages: (a). SimScape Simulation (b). Datasheet [21]**

# IV. Proposed power cycling test platform

One of important tool for reliability testing which have been widely used to do power electronic modules lifetime assessment and robustness evaluation is power cycling (PC) test. Main objective of PC test set up is to stress the device under a repeated thermal cycling inducing the temperature swing $\Delta T_j$ at device junction. The PC tests are designed based on thermal fatigue of devices due to periodic succession of self-heating and cooling phases. Usually, PC tests are performed at low voltage constant current conditions. The device blocking voltage and amplitude of the load current is kept below the nominal rating of the power module. With this approach, a similar stresses and operating condition that a power module would experience in real life application can be exerted under accelerated conditions.

A typical test bench for PC test is shown in Fig. 3 which power module is always in ON-state and a constant current periodically is passing through the die by controlling an external switches. This PC platform is only consider conduction losses of the power module and switching losses are not included to calculate total losses of the switch. Therefore, junction temperature calculation is not so realistic.

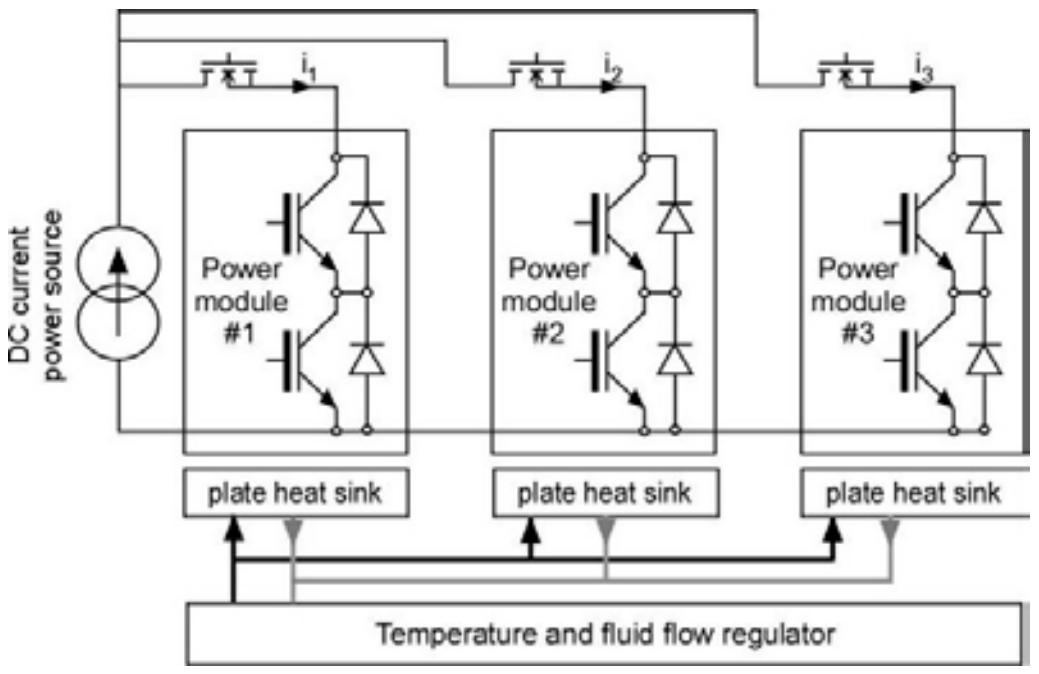

**Fig. 3. A typical test bench for PC test of power switches [22]**

In this paper, the proposed PC test platform, which is shown in Fig. 4.a, is consist of a constant DC voltage supply and couple of power SiC-MOSFETs as DUTs. Each DUT is connected in series with a resistor to generate a constant load current. Based on generated gate pulses for each DUT, the pulsating load current will go through DUT and heat it up. The preliminary experimental test setup for the proposed PC platform is shown in Fig. 4.b. This setup consists of two parallel connected SiC-MOSFETs and will be used to capture critical parameters of power switches. The experimental results will be published in the future.

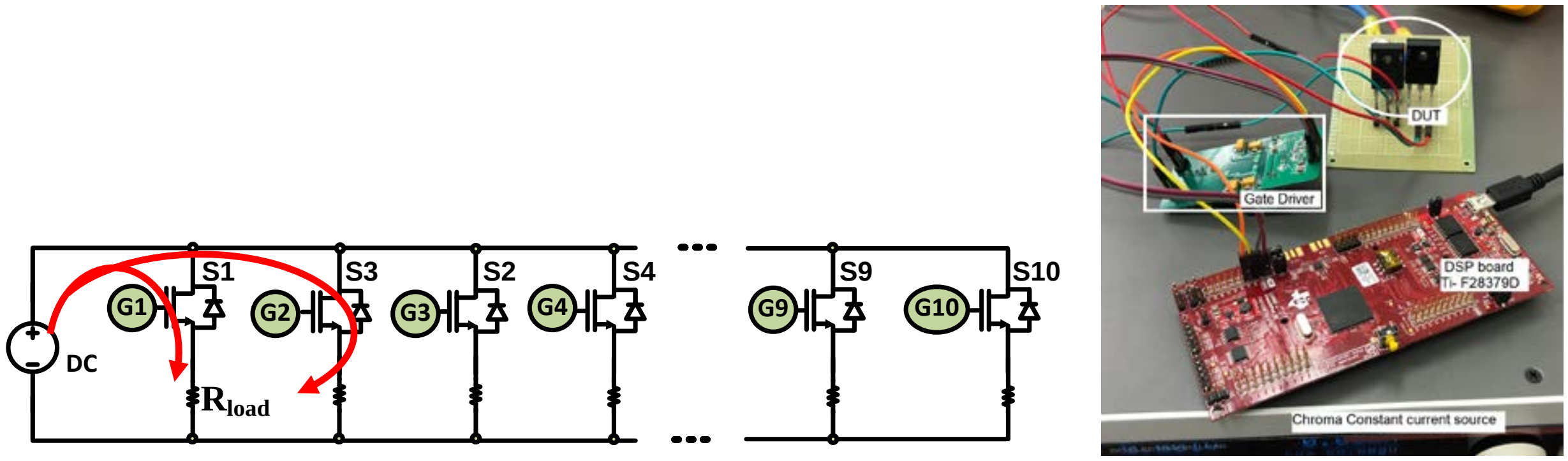

(a) (b)

**Fig. 4. The proposed PC test platform: (a). Schematic (b). Preliminary hardware setup**

Desired temperature swing $\Delta T_j$ at the junction can be obtained by controlling the amplitude and fundamental frequency of the load current cycled through the power switches. The gate pulses of each switch is shown in Fig. 5. In this pattern, during ON-time pulse, the junction will be heated up and at the end of the switching cycle, the junction of SiC-MOSFET will reach its maximum amplitude. As it is shown, switches 1 and 2 are complementary. At the end of switching cycle when switch-1 is not receiving gate pulses anymore and the cooling down process is begun to reduce $T_{J1}$ to ambient temperature, switch 2 is tolerating accelerant power cycle. These switching cycles will be repeated to all other SiC-MOSFETs.

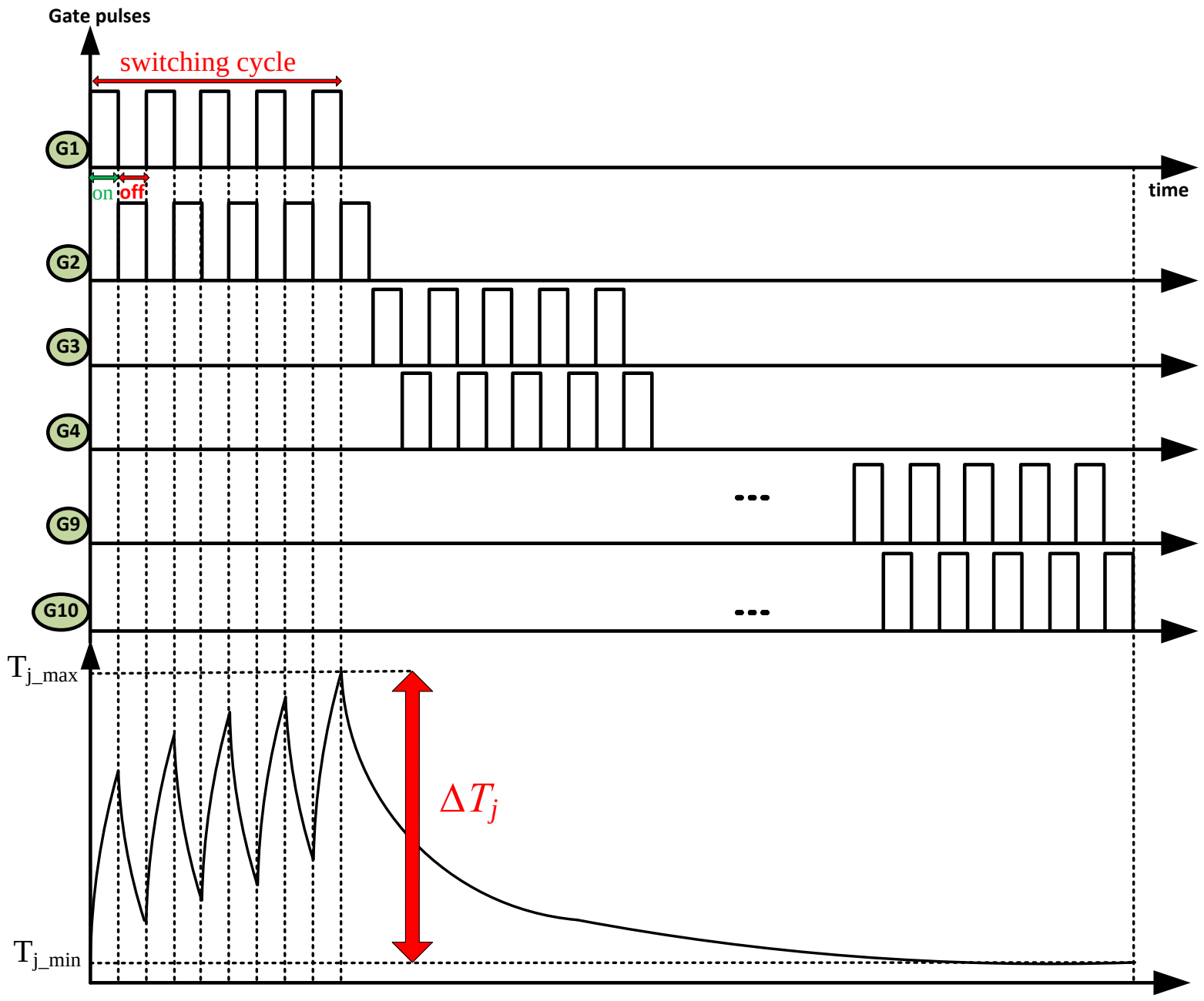

**Fig. 5. The gate pulses of each switch with $\Delta T_j$ sweep in each switching cycle**

Since DC power supply and load current are both constant, all DUTs are dissipating a constant conduction losses ($P_{conduction}$) during ON-state operation. The switching losses ($P_{switching}$) of SiC-MOSFETs are also measured experimentally by applying Double-Pulse-Test (DPT) on the power switches. The DPT setup is shown in Fig. 6.

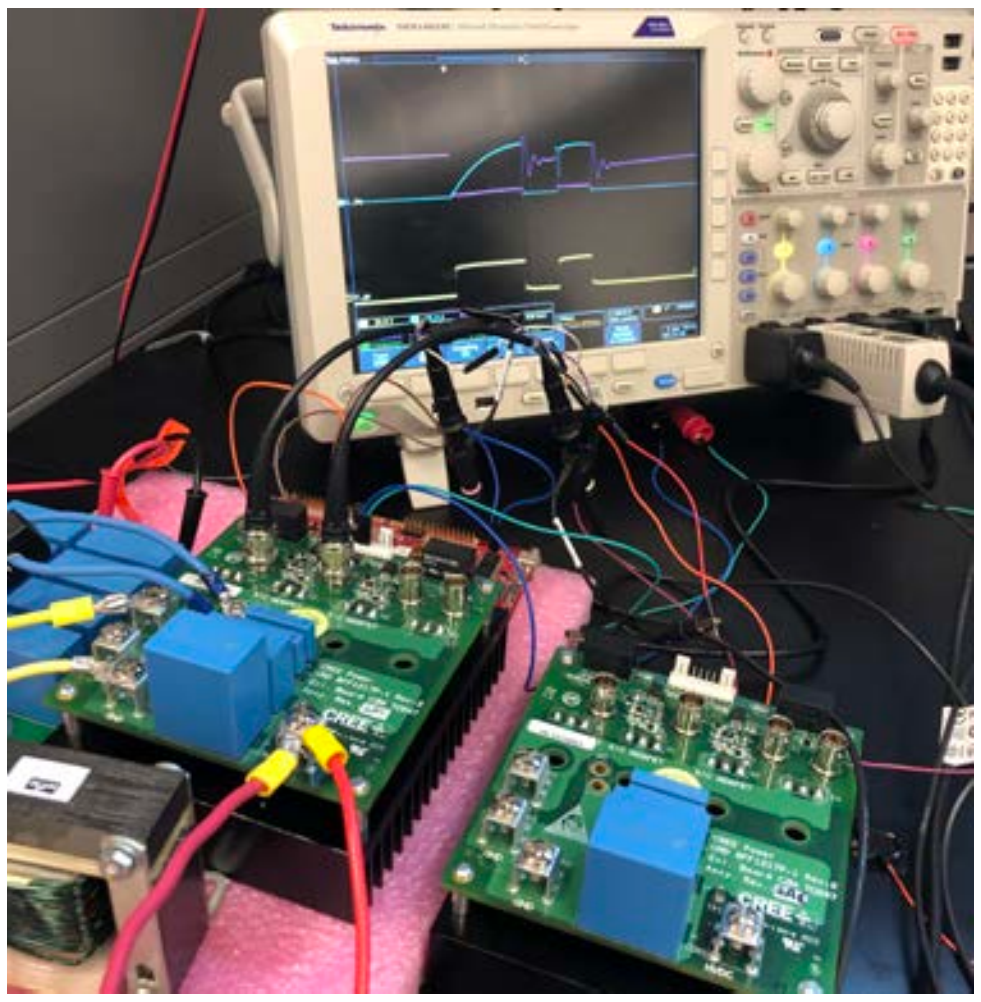

**Fig. 6. Double-Pulse-Test setup**

By knowing $P_{conduction}$, $P_{switching}$, case temperature ($T_C$) and thermal resistance between junction and ambient ($R_{\theta jA}$), the junction temperature ($T_j$) of SiC-MOSFET can be calculated by:

$$T_j = T_c + (P_{cond} + P_{switch}) * R_{\theta JA} \quad (6)$$

The junction temperature variation of switch-1 in one switching cycle is shown in Fig. 7. This data is captured in case temperature at 150°*C*. It is obvious that working in downhole high-temperature ambient will require precise calculation of determining current amplitude and pulse duration to make sure that junction temperature of SiC-MOSFET will not exceed the $T_{j_max}$ limit. It should be mentioned that exceeding $T_{j_max}$ limit will result in over-heating of power switch, which is not included to degrading of SiC-MOSFET, and cannot be considered for RUL prediction. Table-1 shows the variation of amplitude of load current for different ambient temperature with constant pulse duration.

**Table 1. The amplitude of pulsating current in different ambient temperature**

| $T_c$ ( °C) | $I_{pulse}$ (A) |
|---|---|
| 25 | 3.75 |
| 50 | 2.75 |
| 75 | 2.25 |
| 100 | 1.75 |
| 125 | 1.25 |
| 150 | 0.75 |
| $V_{DC}$ = 50 V; $T_{j_max}$ = 175 °C | |

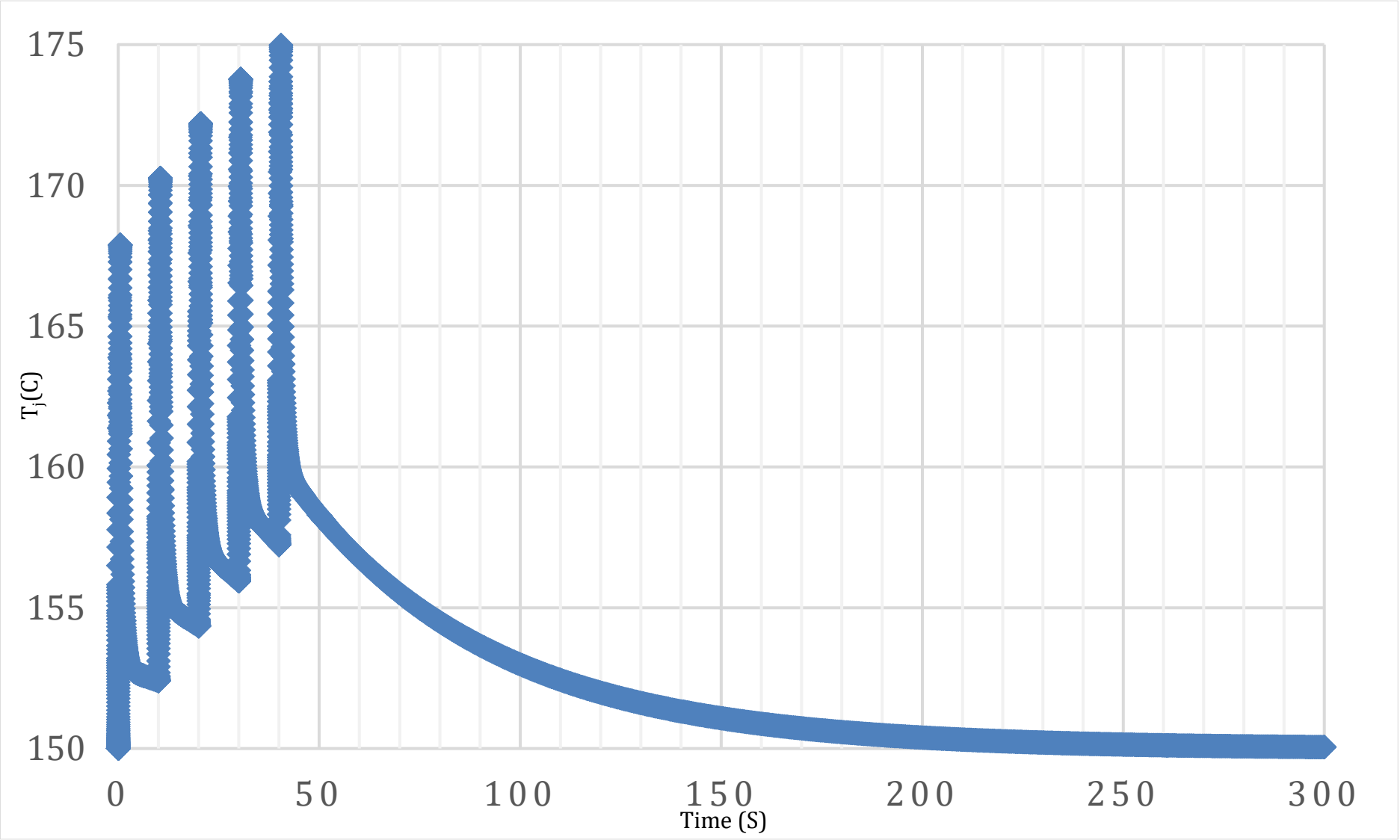

**Fig. 7. The junction temperature variation of switch-1 in one switching cycle**

**Table 2: List of component for $V_{DS(on)}$ measurement circuit**

| Component | Parameter | Part Number | Value |
|---|---|---|---|
| NPN Transistor | Q | Q2N2222A | * |
| Diode | D1 & D2 | 1N4148 | * |
| Resistor | RC | * | 3 KΩ |
| Resistor | RS | * | 1 KΩ |
| Resistor | RB | * | 250 Ω |
| Resistor | RE1 | * | 50 Ω |
| Resistor | RE2 | * | 15 KΩ |
| Capacitor | C1 | * | 1 nF |
| Capacitor | C2 | * | 10 pF |

# V. ON-state Resistance

$R_{DS\ (on)}$ is one of the important parameters related to power MOSFET reliability. During aging tests the on state voltage and drain current are monitored in order to calculate $R_{DS\ (on)}$. In this paper a new $V_{DS(on)}$ circuit (Fig. 8) is used to monitor and measure the changes in the amplitude of drain-source voltage when switch is ON. Table-2 shows the parameters value in $V_{DS(on)}$ measurement circuit.

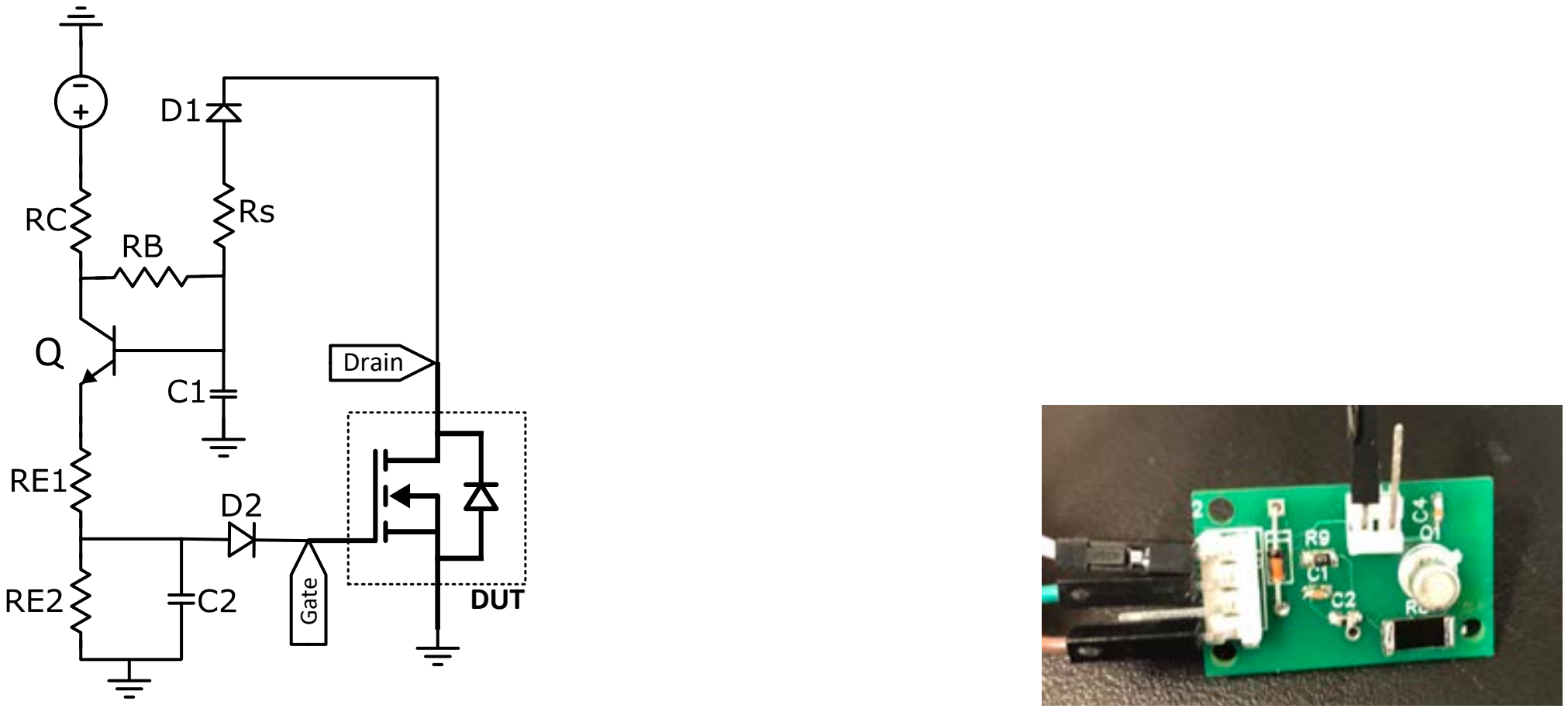

**(a)** **(b)**

**Fig. 8. $V_{DS(on)}$ measurement circuit: (a) Schematic (b). PCB**

Knowing $V_{DS(on)}$ and drain current ($I_D$) which here is pulsating load current, $R_{DS\ (on)}$ can be calculated. The variation in $R_{DS\ (on)}$ during aging test is recorded for the purpose to study the device degradation, which is shown in Fig. 9. During the initial aging cycles, there is negligible variation in $R_{DS\ (on)}$. It is clear that as the number of cycles are increasing during PC test, the $R_{DS\ (on)}$ also varies in increased manner accelerating the degradation of the device.

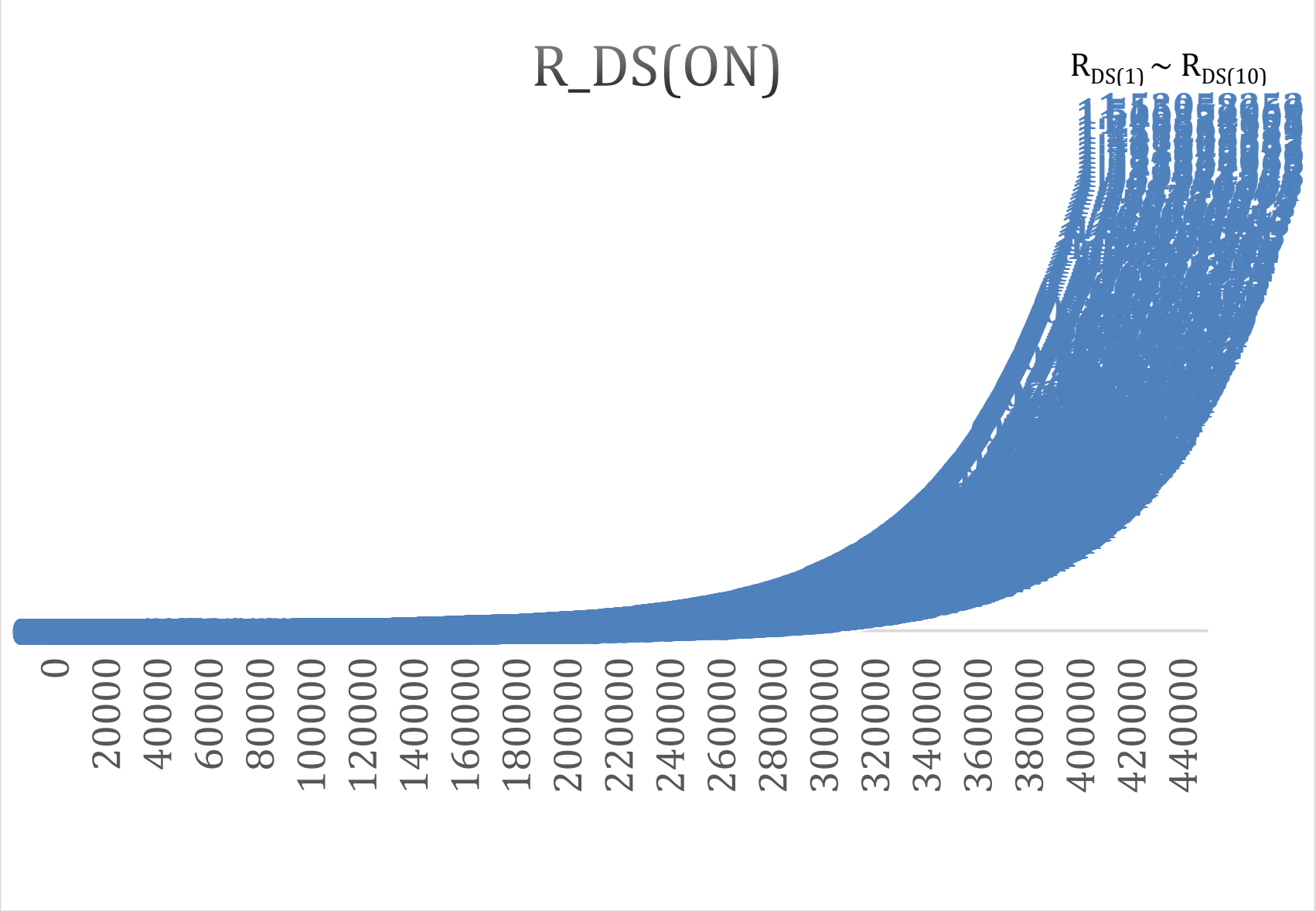

*Fig. 9. $R_{DS(on)}$ variation in PC test cycling*

## VI. Conclusion:

In this paper, a simulation-based modeling of SiC power MOSFETs is used to implement a power cycling (PC) test bench for reliability assessment of SiC-MOSFETs working in harsh environment applications like downhole. The simulation results of SimScape is compared and verified by manufacturer's datasheet. Having more precise and valid simulation-based PC test platform is required for high temperature environment to evaluate testing procedure prior to implement in hardware setup to reduce significant test costs. Besides, a new $V_{DS(on)}$ measurement circuit is introduced to monitor the variation of $R_{DS(on)}$ during the degrading life cycle. The captured data will be used to evaluate RUL prediction of component in a component-level reliability assessment.